\pdfoutput=1

\documentclass[11pt,twoside,a4paper,cmspaper,final,collab]{cms-tdr}
\def\svnVersion{428531P}\def\cmsCernNoTag{CERN-EP-2017-112}\def\cmsCernDate{\today}\def\cmsMessage{Published in the Journal of High Energy Physics as \href{http://dx.doi.org/10.1007/JHEP10(2017)006}{\doi{10.1007/JHEP10(2017)006.}}}

\begin{document}\cmsNoteHeader{TOP-14-008}

\hyphenation{had-ron-i-za-tion}
\hyphenation{cal-or-i-me-ter}
\hyphenation{de-vices}
\RCS$Revision: 428522 $
\RCS$HeadURL: svn+ssh://svn.cern.ch/reps/tdr2/papers/TOP-14-008/trunk/TOP-14-008.tex $
\RCS$Id: TOP-14-008.tex 428522 2017-10-09 16:20:16Z dnoonan $

\providecommand{\statsyst}{\ensuremath{\,(\text{stat+syst})}\xspace}
\providecommand{\cPV}{\cmsSymbolFace{V}\xspace}
\newcommand{\ejets}{\ensuremath{\Pe\text{+jets}}\xspace}
\newcommand{\mujets}{\ensuremath{\Pgm\text{+jets}}\xspace}
\newcommand{\sigmaietaieta}{\ensuremath{\sigma_{\eta\eta}}\xspace}
\newcommand{\ttgamma}{\ensuremath{\ttbar \text{+} \gamma}\xspace}
\newcommand{\TTJets}{\ensuremath{\ttbar\text{+jets}}\xspace}
\newcommand{\ZJets}{\ensuremath{\Z\text{+jets}}\xspace}
\newcommand{\Wgamma}{\ensuremath{\PW\text{+}\gamma}\xspace}
\newcommand{\Zgamma}{\ensuremath{\Z\text{+}\gamma}\xspace}
\newcommand{\Vgamma}{\ensuremath{\cPV\text{+}\gamma}\xspace}
\newcommand{\WJets}{\ensuremath{\PW\text{+jets}}\xspace}

\newcommand{\VJets}{\ensuremath{\cPV\text{+jets}}\xspace}
\newcommand{\Mthree}{\ensuremath{M_3}\xspace}
\newcommand{\ttgammaSF}{\ensuremath{\mathrm{SF}_{\ttgamma}}\xspace}
\newcommand{\VgammaSF}{\ensuremath{\mathrm{SF}_{\Vgamma}}\xspace}
\newcommand{\jetToPhotonSF}{\ensuremath{\mathrm{SF}_{\text{jet}\to \gamma}}\xspace}

\newcolumntype{P}[1]{D{~}{\,\pm\,}{#1}}
\newcommand\hd[1]{\multicolumn{1}{c}{#1}}
\newcommand\hdd[1]{\multicolumn{1}{c|}{#1}}

\newcommand{ \muJetsExtrapolatedCX }{ \ensuremath{ 453 } }
\newcommand{ \muJetsExtrapolatedCXErr }{ \ensuremath{ 124 } }
\newcommand{ \muJetsFidRatioWithError }{ \ensuremath{ (4.7 \pm 1.3)\times 10^{-4} } }
\newcommand{ \muJetsFidCrossSecFB }{ \ensuremath{ 115 } }
\newcommand{ \muJetsFidCrossSecFBErr }{ \ensuremath{ 32 } }
\newcommand{ \muJetsFidEff }{ \ensuremath{ 0.1268 } }
\newcommand{ \muJetsFidEffErr }{ \ensuremath{ 0.0070 } }
\newcommand{ \muJetsKinAcc }{ \ensuremath{ 0.2551 } }
\newcommand{ \muJetsKinAccErr }{ \ensuremath{ 0.0014 } }
\newcommand{ \muJetsTTbarTopPresel }{ \ensuremath{ 219128 } }
\newcommand{ \muJetsTTbarTopPreselErr }{ \ensuremath{ 1869 } }
\newcommand{ \muJetsTTbarTopEffRounded }{ \ensuremath{ 0.046 } }
\newcommand{ \muJetsTopPurity }{ \ensuremath{ 0.68 } }
\newcommand{ \muJetsTopPurityErr }{ \ensuremath{ 0.06 } }
\newcommand{ \muJetsPhoPurity }{ \ensuremath{ 0.53 } }
\newcommand{ \muJetsPhoPurityErr }{ \ensuremath{ 0.06 } }
\newcommand{ \eGammaSF }{ \ensuremath{ 1.46 } }
\newcommand{ \eGammaSFErr }{ \ensuremath{ 0.20 } }
\newcommand{ \eJetsExtrapolatedCX }{ \ensuremath{ 582 } }
\newcommand{ \eJetsExtrapolatedCXErr }{ \ensuremath{ 187 } }
\newcommand{ \eJetsFidRatioWithError }{ \ensuremath{ (5.7 \pm 1.8)\times 10^{-4} } }
\newcommand{ \eJetsFidCrossSecFB }{ \ensuremath{ 138 } }
\newcommand{ \eJetsFidCrossSecFBErr }{ \ensuremath{ 45 } }
\newcommand{ \eJetsFidEff }{ \ensuremath{ 0.1198 } }
\newcommand{ \eJetsFidEffErr }{ \ensuremath{ 0.0071 } }
\newcommand{ \eJetsKinAcc }{ \ensuremath{ 0.2380 } }
\newcommand{ \eJetsKinAccErr }{ \ensuremath{ 0.0014 } }
\newcommand{ \eJetsTTbarTopPresel }{ \ensuremath{ 162168 } }
\newcommand{ \eJetsTTbarTopPreselErr }{ \ensuremath{ 1565 } }
\newcommand{ \eJetsTTbarTopEffRounded }{ \ensuremath{ 0.034 } }
\newcommand{ \eJetsTopPurity }{ \ensuremath{ 0.70 } }
\newcommand{ \eJetsTopPurityErr }{ \ensuremath{ 0.08 } }
\newcommand{ \eJetsPhoPurity }{ \ensuremath{ 0.57 } }
\newcommand{ \eJetsPhoPurityErr }{ \ensuremath{ 0.06 } }
\newcommand{ \combinedFidRatioWithErrorPerChannel }{ \ensuremath{ (5.2 \pm 1.1)\times 10^{-4} } }
\newcommand{ \combinedFidCrossSecPerChannelFB }{ \ensuremath{ 127 } }
\newcommand{ \combinedFidCrossSecErrPerChannelFB }{ \ensuremath{ 27 } }
\newcommand{ \combinedExtrapolatedCXPerChannel }{ \ensuremath{ 515 } }
\newcommand{ \combinedExtrapolatedCXPerChannelErr }{ \ensuremath{ 108 } }

\cmsNoteHeader{TOP-14-008}
\title{Measurement of the semileptonic $\ttbar+\gamma$ production cross section in pp collisions at $\sqrt{s}=8$\TeV}

\date{\today}

\abstract{
A measurement of the cross section for top quark-antiquark ($\ttbar$) pairs produced in association with a photon in proton-proton collisions at $\sqrt{s}=8$\TeV is presented.
The analysis uses data collected with the CMS detector at the LHC, corresponding to an integrated luminosity of 19.7\fbinv.
The signal is defined as the production of a $\ttbar$ pair in association with a photon having a transverse energy larger than 25\GeV and an absolute pseudorapidity smaller than 1.44.
The measurement is performed in the fiducial phase space corresponding to the semileptonic decay chain of the $\ttbar$ pair, and the cross section is measured relative to the inclusive $\ttbar$ pair production cross section.
The fiducial cross section for associated $\ttbar$ pair and photon production is found to be $127 \pm 27\statsyst$\unit{fb} per semileptonic final state.
The measured value is in agreement with the theoretical prediction.
}

\hypersetup{
pdfauthor={CMS Collaboration},
pdftitle={Measurement of the semileptonic ttbar gamma production cross section in pp collisions at sqrt(s) = 8 TeV},
pdfsubject={CMS},
pdfkeywords={CMS, physics, top quark}}

\maketitle

\section{Introduction}\label{sec:intro}

As the heaviest elementary particle in the standard model (SM), the top quark has the potential to provide insights into physics beyond the SM (BSM).
Many BSM models introduce changes within the top quark sector~\cite{Han:2008xb, Bernreuther:2008ju}, which can be constrained by precise measurements of the cross sections and properties of top quark production channels ~\cite{Buckley:2015lku}.
By measuring the associated production cross section of a top quark-antiquark pair and a photon (\ensuremath{\ttbar \text{+} \gamma}),
the coupling of the top quark and the photon is probed~\cite{Baur1, Bouzas1}.  Any deviation
of the measured cross section value from the SM prediction would be an indication of BSM physics, such as
the production of an exotic quark with electric charge of $-$4/3,
or a top quark with an anomalous electric dipole moment~\cite{Bylund:2016phk,Schulze:2016qas}.

\begin{figure}[b]
  \centering
    \includegraphics[width=0.3\textwidth]{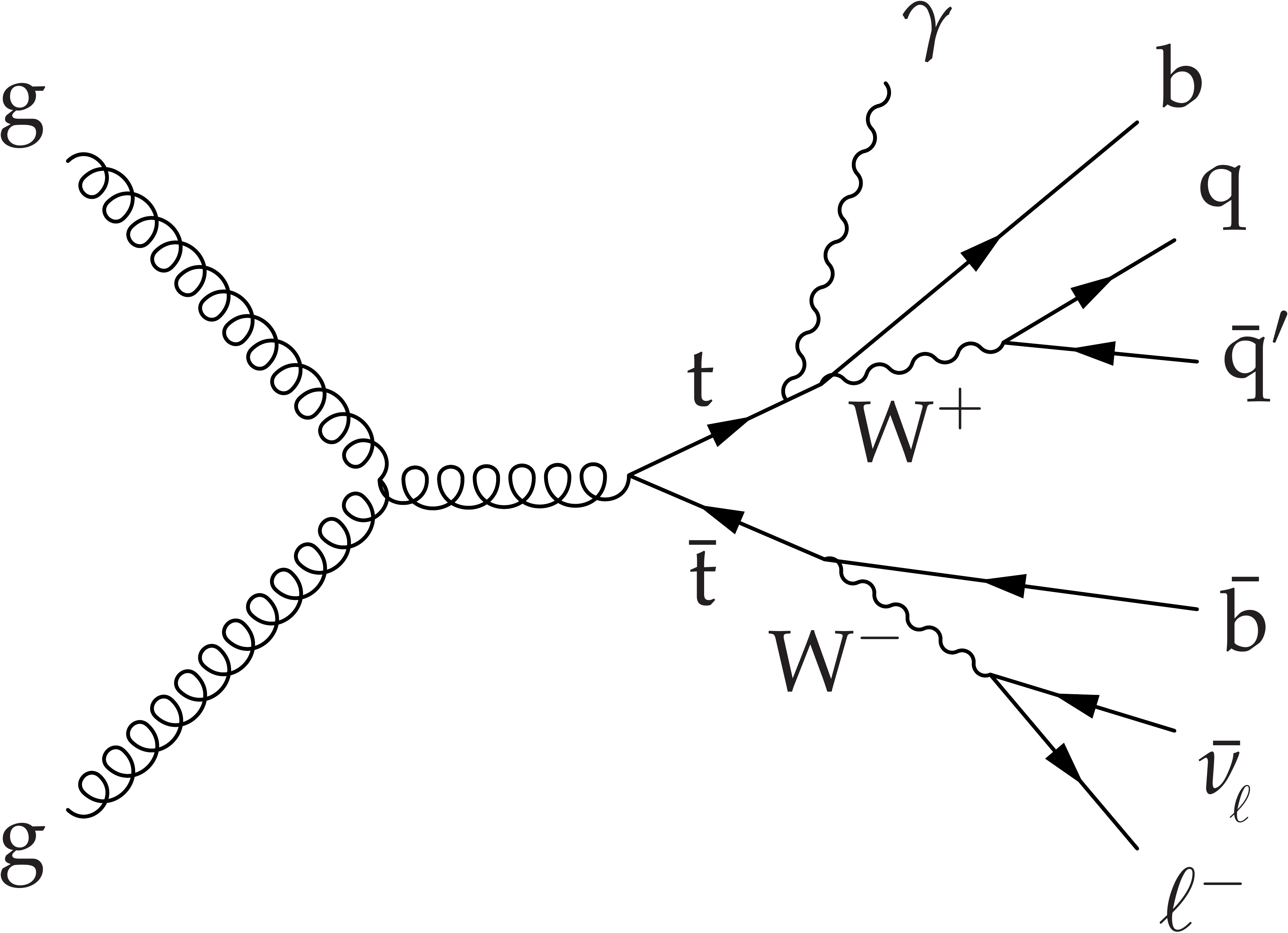}
    \hspace{0.1\textwidth}
    \includegraphics[width=0.3\textwidth]{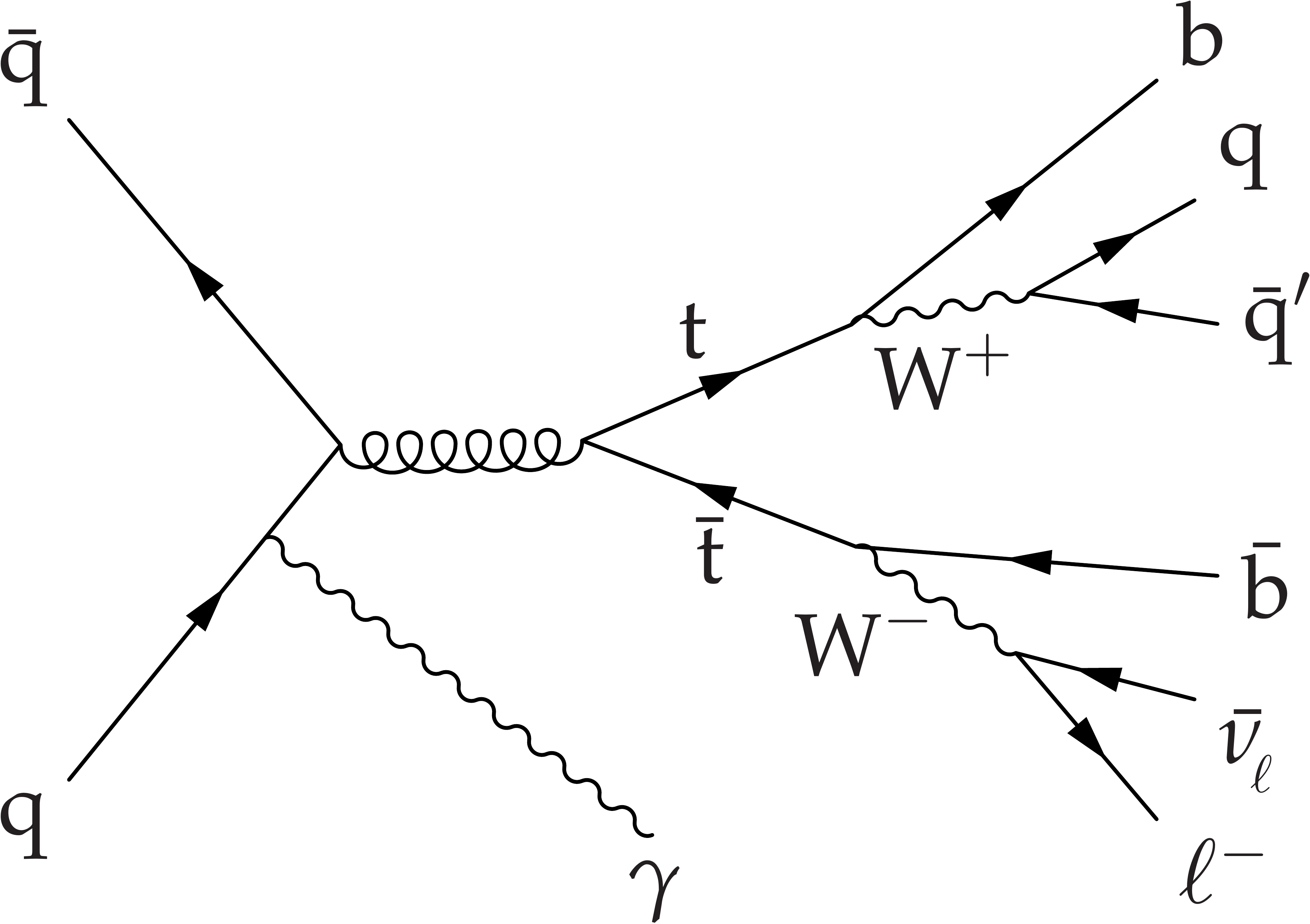}
    \caption{Dominant Feynman diagrams for the \ttgamma signal process in the semileptonic final state where the \ttbar pair is produced through gluon-gluon fusion with a photon emitted from one of the top quarks (left), and through quark-antiquark annihilation with a photon emitted from one of the initial partons (right).}
    \label{fig:FeynDiagramSemilep}

\end{figure}

As the top quark predominantly decays to a W boson and a b quark, the \ttgamma
production can be identified by the presence of a photon candidate
 and the decay products of a pair of top quarks, namely two
jets from the hadronization of two b quarks, and the decay products of
a pair of W bosons.
In this analysis, events are selected in which one W boson decays
leptonically, resulting in an electron or muon and a corresponding
neutrino $\nu$, and the other W boson decays hadronically.
Examples of two Feynman diagrams for the \ttgamma process in the semileptonic final states are shown in Fig. \ref{fig:FeynDiagramSemilep}.
In the signal definition we include possible contributions from $\PW\to\tau\nu_\tau$, where the $\tau$ lepton decays further into an electron or a muon.
The presence of a charged lepton from the W boson decay
significantly improves the power to reject dominant
backgrounds from multijet processes and allows for efficient triggering of
signal events using single-lepton triggers.

Measurements of the production cross section of \ttgamma have been performed by the CDF Collaboration at the
Tevatron using $\Pp\PAp$ collisions at $\sqrt{s} = 1.96\TeV$~\cite{CDFttgamma}, and by the
ATLAS Collaboration at the LHC using  $\Pp\Pp$ collisions at $\sqrt{s} = 7\TeV$~\cite{ATLASttgamma7TeV} and $\sqrt{s} = 8\TeV$~\cite{ATLASttgamma8TeV}.
These results are in agreement with the SM predictions within uncertainties~\cite{TevatronTTgammaPrediction}.

In this paper, the measurement of the \ttgamma production cross section in $\Pp\Pp$ collisions at $\sqrt{s}=8\TeV$ is presented.
The analysis is based on a data sample corresponding to an integrated luminosity of $19.7\fbinv$, recorded with the CMS detector in 2012.
The measurement of the \ttgamma production cross section in the semileptonic decay channel is performed relative to the \ttbar production cross section.
The \ttgamma cross section is measured in a fiducial kinematic region defined by the presence of exactly one charged lepton and corresponding neutrino, at least three jets, and a photon within the selection requirements.

\section{The CMS detector}

The central feature of the CMS apparatus is a superconducting solenoid of 6\unit{m} internal diameter, providing a magnetic field of 3.8\unit{T}.
Within the solenoid volume are a silicon pixel and strip tracker, a lead tungstate crystal electromagnetic calorimeter (ECAL),
and a brass and scintillator hadron calorimeter, each composed of a barrel and two endcap sections. Forward calorimeters extend the
pseudorapidity $\eta$ coverage provided by the barrel and endcap detectors.
Muons are detected in gas-ionization chambers embedded in the steel flux-return yoke outside the solenoid.
In the barrel section of the ECAL, an energy resolution of about 1\% is achieved for unconverted or late-converting photons in the tens of\GeV energy range, relevant to this analysis. The remaining barrel photons have a resolution of about 1.3\% up to $\abs{\eta} = 1$, rising to about 2.5\% at $\abs{\eta} = 1.4$. In the endcaps, the resolution of unconverted or late-converting photons is about 2.5\%, while the remaining endcap photons have a resolution between 3 and 4\%~\cite{CMS:EGM-14-001}.
A more detailed description of the CMS detector, together with a definition of the coordinate system used and the relevant kinematic variables, can be found in Ref.~\cite{Chatrchyan:2008zzk}.

\section{Signal and background modeling}

The signal process produces events in which a pair of top quarks is produced in association with a photon.
This process includes photons radiated from the top quarks as well as from initial state partons or the decay products of the top quarks.
The simulation of the \ttgamma signal process is performed in the region with photons having transverse momentum (\pt) of at least 13\GeV and $\abs{\eta}<3.0$, as well as having a separation from all other generated particles of at least $\Delta R > 0.3$, where $\Delta R = \sqrt{\smash[b]{(\Delta\phi)^2 + (\Delta\eta)^2}}$, and $\Delta\phi$ and $\Delta\eta$ are the differences in the azimuthal angle (in radians) and pseudorapidity, respectively, between the generated particles and the photon.
For the purpose of this analysis, nonprompt photons originating from jets are not included in the definition of the \ttgamma signal process.

The $\ttgamma$ signal process is simulated at leading order (LO) using
the \MADGRAPH v5.1.3.30 generator~\cite{madgraph5}.
The dominant backgrounds, $\ttbar$, \VJets, and \Vgamma
(where $\cPV=\PW$, $\cPZ$), are also simulated using the \MADGRAPH generator.
Single top quark production is simulated at next-to-leading order (NLO)
using the \POWHEG v1.0 r1380 event generator~\cite{POWHEG, POWHEG2, POWHEG:tW, POWHEG:SingleTop}.
In order to avoid any overlap between the simulation of the \ttgamma signal and the inclusive \ttbar process,
events that fall under the \ttgamma signal definition are removed from \ttbar simulation.
Overlap between \Vgamma and \VJets simulation is also taken into account by removing events from
\VJets samples, which are accounted for in the \Vgamma simulation.
Approximately 1\% of events from \ttbar simulation and approximately 3\% of \VJets events are removed through this procedure.

The parton showering and hadronization for all simulated samples are
handled by \PYTHIA v6.426~\cite{Pythia}, with the decays of $\tau$ leptons
modeled with \TAUOLA v27.121.5~\cite{tauola}.
The CTEQ6L1 and CTEQ6M~\cite{cteq} parton distribution functions (PDFs)
are used for samples simulated at LO and NLO, respectively.
A top quark mass $m_\cPqt=172.5\GeV$ is used in the simulation.
The response of the full CMS detector is simulated with \GEANTfour v9.4~\cite{GEANT4, GEANT4_2nd},
followed by a detailed trigger simulation and event reconstruction.
The \PYTHIA event generator is used to simulate the presence of additional $\Pp\Pp$ interactions
in the same and nearby bunch crossings (``pileup''). Simulated events are reweighted to correct
for differences between the number of pileup interactions observed from data and
the number produced in the simulation.

A cross section of $244.9\pm1.4\stat^{+6.3}_{-5.5}\syst\pm6.4\lum$\unit{pb} is used to normalize the \ttbar background~\cite{cmsTTbarCXDilep2016}.
The next-to-next-to-leading-order (NNLO) SM prediction is calculated with \textsc{fewz} v3.1~\cite{Melnikov:2006a,Melnikov:2006b}
for the \VJets backgrounds.
The \Wgamma and \Zgamma simulations are normalized to their NLO predictions, calculated with \MCFM v6.6~\cite{Campbell:2010ff}.  Values of 553.9\unit{pb} for the leptonic decay of the W+gamma process and 159.1\unit{pb} for the leptonic decay of the Z+gamma process are used.
The single top quark samples are normalized to their approximate NNLO predictions~\cite{Kidonakis:2010a,Kidonakis:2010b}.

\section{Event reconstruction and selection} \label{sec:EventReconstruction}

The final state of the signal process in the semileptonic decay channel consists
of a high-\pt charged lepton, momentum imbalance due to the presence of a
neutrino, jets originating from both the $\cPqb$ quarks and from the decay of a W boson, and
an energetic photon.
Events with either a high-\pt electron or muon are initially selected through a single-lepton trigger.
Events in the \ejets final state must pass a trigger requiring an electron with $\pt > 27$\GeV within $\abs{\eta}<2.5$
and a relative isolation of less than 0.2, where the relative isolation is defined as the sum of the \pt of all particles, excluding the lepton, within a cone around the lepton of $\Delta R  = 0.3$, divided by the \pt of the lepton.
The \mujets final state requires a single-muon trigger selecting a muon with $\pt>24$\GeV within $\abs{\eta}<2.1$ and relative isolation less than 0.3 within $\Delta R =0.4$.
Events are additionally required to have a well reconstructed primary vertex~\cite{Chatrchyan:2014fea}, chosen as the one having the largest sum $\pt^2$ of the tracks associated with it.

The particle-flow (PF) algorithm is used to reconstruct individual particles
in the event~\cite{particleFlowPaper}. The PF objects include
electrons, muons, charged and neutral hadrons, photons, and an imbalance of the
transverse momentum. The following describes the selection of reconstructed
objects that are used in the analysis.

Electrons are reconstructed from energy deposits in the ECAL matched to a
track from the tracker~\cite{electronReconstruction}.
Electrons are required to have $\pt > 35$\GeV and $\abs{\eta} < 2.5 $.
excluding the transition region between the barrel and endcap of the ECAL, $1.44 < \abs{\eta} < 1.57$.
Electrons from the decay of the top quark are expected to be isolated from other activity in the detector
and thus have a requirement that the relative isolation must be less than 0.1.
Selected electrons are required to be originating from the primary vertex, and are rejected if
identified as likely having originated from a converted photon.
Additionally, a multivariate-based identification is applied to reduce the contribution
from nonprompt or misidentified electrons.
Electrons that fail the above criteria, but pass looser identification requirements ($\pt>20\GeV$, $\abs{\eta} < 2.5 $, and a relative isolation less than 0.2 within a cone of size $\Delta R  = 0.3$) are considered to be ``loose'' electrons.
The presence of loose electrons can then be used to reject events from the dilepton final state.

Muons are reconstructed based on measurements from both the tracker and
muon systems. Selected muons are required to have $\pt > 26\GeV$ and
$\abs{\eta} < 2.1$. A requirement on relative isolation less than 0.2
within a cone of $\Delta R = 0.4$ is applied.
Loose muons are defined as failing  the tight requirements but passing a selection in which the \pt threshold is
lowered to 10\GeV and $\abs{\eta} < 2.5$, with the same requirement on the relative isolation as the tight selection.

Jets are reconstructed from PF candidates clustered using the
anti-$\kt$ algorithm with a distance parameter
of 0.5~\cite{Cacciari:2008gp, FastJet}. Jets must have $\pt> 30\GeV$
and $\abs{\eta}<2.4$.
To remove the contribution to the jet energy from pileup interaction, charged hadrons candidates associated with other vertices are not included in the clustering, and an offset correction to the energy is applied for the contribution of neutral hadrons that would fall within the jet area.
Additionally, corrections for the jet energy scale and resolution are applied in simulation, to account for imperfect measurements of the
energy of the jet in the detector~\cite{JES}.

Jets are identified as originating from the hadronization of b quarks (b tagged) using
the combined secondary vertex algorithm, which combines secondary
vertex and track-based lifetime information to provide a discriminant
between jets originating from the fragmentation of b quarks
and light quarks or gluons.
The b tagging algorithm has an efficiency of approximately 70\%, while having a probability of incorrectly b tagging a light jet of only 1.4\%~\cite{Btagging, BTV12001}.

Photons are reconstructed as energy deposits in the ECAL that are not matched to track seeds in the pixel detector~\cite{CMS:EGM-14-001}.
The photon is required to have $\pt>25\GeV$ and $\abs{\eta}<1.44$ (ECAL barrel).
A selection based on the shape of the shower caused by the photon in the ECAL is applied using the $\sigmaietaieta$ variable, which measures the lateral spread of energy in the $\eta$ space~\cite{CMS:EGM-14-001}.
Selected photons are required to have $\sigmaietaieta<0.012$.
This is used to distinguish genuine photons from hadronic activity that can be reconstructed as a photon, as the latter will tend to produce a wider energy spread in $\eta$, leading to a larger value of \sigmaietaieta.
As photons can convert into a pair of electrons before reaching the calorimeter, photon showers along $\phi$ can be larger compared to that of an electron.
Thus, the isolation is defined differently for photons than it was for leptons,
in order to account for a possible energy leakage along $\phi$.
A characteristic photon energy deposition profile, or ``footprint'',
is used to restrict the area used to calculate isolation of the photon
candidate.
The charged-hadron isolation variable for photons is defined as the sum of the \pt of all charged hadrons spatially separated from the photon candidate by $\Delta R = 0.3$, but not falling within the photon footprint.
The charged-hadron isolation is required not to exceed 5\GeV for selected photons, to help distinguish prompt photons from nonprompt photons produced from hadronic activity.

The missing transverse momentum (\ptmiss) is defined as the magnitude of the vector sum  of the momenta of all reconstructed PF candidates in the event, projected on the plane perpendicular to the beams.

The final event selection is divided in two steps:
a preselection designed to select events with the
same topology as top quark pairs (referred to as the ``top quark selection''),
and a ``photon selection''.
The top quark selection requires:
\begin{itemize}
\item exactly one lepton passing the selection requirements (either an electron or muon);
\item no other lepton candidates passing loose selection criteria;
\item at least 3 jets, with at least one of these jets passing the b tagging requirement; and
\item $\ptmiss > 20\GeV$.
\end{itemize}
The photon selection requires that events pass the top quark selection and additionally
have at least one photon passing the identification and isolation requirements described above.

\section{Analysis strategy}

After the photon selection is applied, over half of the events in simulation originate from background processes, and not \ttgamma production.
The two largest backgrounds are from \ttbar events that have a nonprompt photon coming from jets in the event and from \Vgamma events.
There is not a single variable that can sufficiently discriminate both of these backgrounds from the \ttgamma signal.
The \Vgamma background can be differentiated from \ttgamma events by attempting to reconstruct a top quark in the event.
However, \ttbar events are very similar to the signal in this respect.
Alternatively, the nonprompt photon from the \ttbar background will tend to be less isolated than the photons from the \ttgamma signal, but the photon isolation variable does not have discrimination power to distinguish the \Vgamma background from \ttgamma events.
In order to be able to distinguish both \ttbar and \Vgamma background events, both of these methods are used and the results are combined to measure the \ttgamma yield observed in data.

The fraction of events passing the photon selection containing top quark pairs, referred to as the ``top quark purity'', can be measured by reconstructing the hadronically decaying top quark in the event.
The \Mthree variable, defined as the invariant mass of the three-jet combination that gives the highest vector sum of individual jet transverse momenta, is used for this purpose.
Section \ref{sec:M3Fit} describes the fit to the distribution of the \Mthree variable, used to distinguish top quark pair events from other backgrounds.

Section \ref{sec:photonPurity} describes the measurement of the ``photon purity'', defined as the fraction of reconstructed photons in the selection region, which come from genuine, isolated photons as opposed to misidentified photons originating from jets.
A fit to the photon isolation is used to measure this quantity, which can discriminate between the genuine photons expected from signal and the nonprompt photons from the \ttbar background.

The fits for extracting the top quark and photon purity are performed sequentially, and then the values are used in a likelihood function, from which a fit is performed to extract the number of events that originate from the \ttgamma signal process.
The likelihood fit and extraction of the number of \ttgamma events are described in Section \ref{sec:likelihoodFit}.

\section{Multijet and Z+jets background estimation} \label{sec:QCDDYestimate}

The quantum chromodynamics (QCD) multijet process is not adequately modeled by simulation, so a data-based approach is applied to measure the shape and normalization of this background component.
The shape of the QCD multijet background is taken from a sideband region in data.
The sideband region is defined by inverting the lepton relative isolation requirement, selecting leptons with a relative isolation greater than 0.25.
Additionally, in the \ejets final state the requirement on the multivariate-based electron identification is inverted, selecting electrons that would typically be identified as misidentified or nonprompt.
This control region is dominated by QCD multijet events, with only minor contributions from other processes such as \ttbar and \WJets.
The small contribution in the control region from other processes is subtracted using simulation to provide shapes of the variable distributions used in the analysis.

The normalization of the QCD multijet background is measured through a binned maximum-likelihood fit to the \ptmiss distribution after the standard top quark selection is applied.
The distribution of \ptmiss is softer in the QCD multijet background than the other processes considered, and thus provides some discriminating power for this background.
For the purposes of the fit, the selection requirement on \ptmiss is removed, in order to improve the discriminating power of the fit by bringing in more multijet events into the fit region.
Two distributions are used in the fit, one for the multijet background and one for the contribution from all other processes.
The distribution for the multijet background is taken from the shape found in the sideband control region, while the second distribution is taken from the sum of all simulated events (which does not include the QCD background component).
The fit is performed separately in the \ejets and \mujets final states, and the results are used to scale the QCD multijet background distributions later in the analysis.

The normalization and modeling of the \ZJets background distribution is taken from simulation, but the normalization is corrected by applying a scale factor derived from a fit to data.
In order to check the normalization, the selection is modified, selecting same-flavor dilepton events, while keeping all other top quark selection requirements in place.
A binned maximum-likelihood fit is performed to the dilepton invariant mass for events passing this modified selection.
The fit is performed using two normalized distributions (templates) from simulation, a \ZJets template and a background template, which predominantly contains \ttbar events.
Scale factors for the normalization of the \ZJets background are derived from the fit and applied to the simulation.

\section{Estimate of top quark pair production} \label{sec:M3Fit}

The number of events containing top quark pairs, both after the top quark selection and for events passing the photon selection,
are extracted through a binned maximum-likelihood fit to the distribution of the \Mthree variable.
In events with semileptonic decays of the top quark pair, the \Mthree variable provides a simple reconstruction of the hadronically decaying top quark, and has a distribution peaking at the mass of the top quark.
Other processes have a wider \Mthree distribution.

Two separate fits are performed to the \Mthree distribution.
The first fit is performed after the top quark selection, to extract the total number of \ttbar events passing the selection, $N_{\ttbar}$.
The second fit is performed for events passing the photon selection in order to measure the top quark purity.

\subsection[Measurement of the top quark-antiquark pair yield]{Measurement of the \ttbar yield}
The fit to the \Mthree distribution for events passing the top quark selection is used to extract the total number of top quark pairs, used for measuring the \ttbar component of the cross section ratio.
The fit uses three templates: associated to top quark events (taken from \ttbar and \ttgamma simulation), \WJets, and other background processes.
The template for the other background processes is a combination of the data-based QCD multijet background and all other simulated samples.
In the fit, the normalizations of the top quark and \WJets templates are allowed to float, while the normalizations of the other backgrounds templates are kept fixed.
The QCD multijet background is normalized to the fit to the \ptmiss distribution, while other simulated samples are scaled to their theoretical cross sections.
From the fit, $\eJetsTTbarTopPresel \pm \eJetsTTbarTopPreselErr\ \text{(stat)}$ and $\muJetsTTbarTopPresel \pm \muJetsTTbarTopPreselErr\ \text{(stat)}$ \ttbar events are observed in the \ejets and \mujets final states, respectively, consistent with the expected total number of \ttbar events.
The fit results are used to scale the normalization of the \ttbar and \WJets contributions in the rest of the analysis.

\subsection{Measurement of the top quark purity}
\begin{figure}
   \centering
     \includegraphics[width=0.6\textwidth]{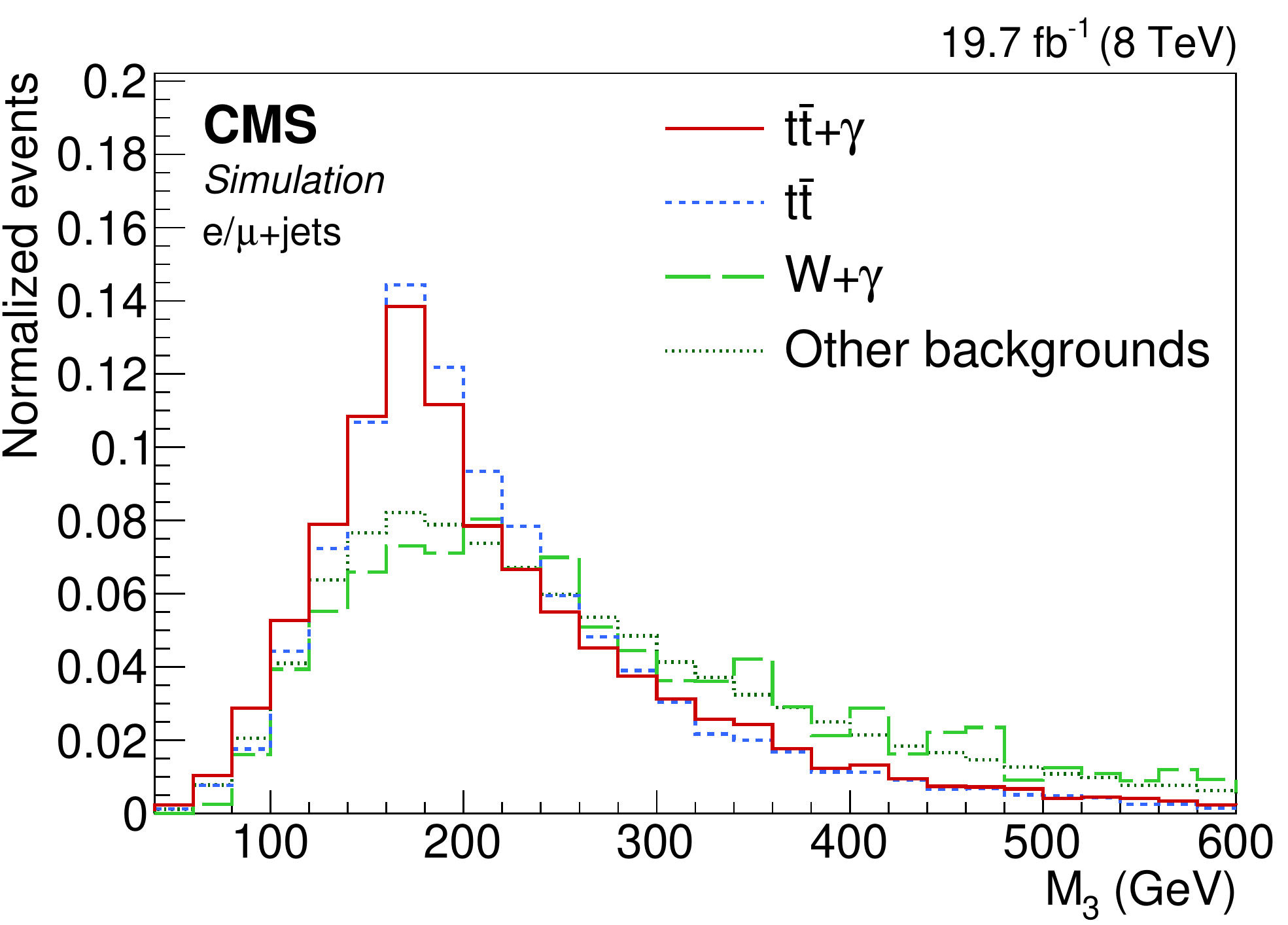}
     \caption{Normalized distributions of the \Mthree variable for \ttgamma, \ttbar, \Wgamma,  and other background processes in a combination of the \ejets and \mujets final state after the photon selection.}
     \label{fig:M3_normalized}

 \end{figure}

After the photon selection, a fit to the \Mthree distribution is used to measure the top quark purity.
The fit uses three templates: associated to top quark events, \Wgamma events, and the sum of all other processes.
In the fit, the normalizations of the top quark and \Wgamma templates are varied, while the templates of all other processes remain fixed.
The top quark template contains simulated events for both \ttgamma and \ttbar samples.
Figure \ref{fig:M3_normalized} shows the normalized \Mthree distributions for \ttgamma, \ttbar, \Wgamma, and other background processes.
The backgrounds from non-top quark processes have a wider distribution in this variable, while the \ttgamma and \ttbar processes peak near the top quark mass with a tail caused by events with an incorrect assignment of the jets.
The relative contributions of the \ttgamma and \ttbar samples to the top quark template are computed from the expected yields from simulation, though this does not change the shape of the top quark template as the two distributions are compatible.
After the photon selection is applied, the distribution of the \Mthree variable in many of the background processes begins to suffer from fluctuations caused by the limited number of simulated events.
Because the photon selection does not change the shape of the \Mthree distribution, the problem is solved by taking the shapes for the non-\ttbar processes from the events after the top quark selection, while retaining the normalization of the samples observed after the photon selection is applied.

\begin{figure}
    \centering
      \includegraphics[width=0.6\textwidth]{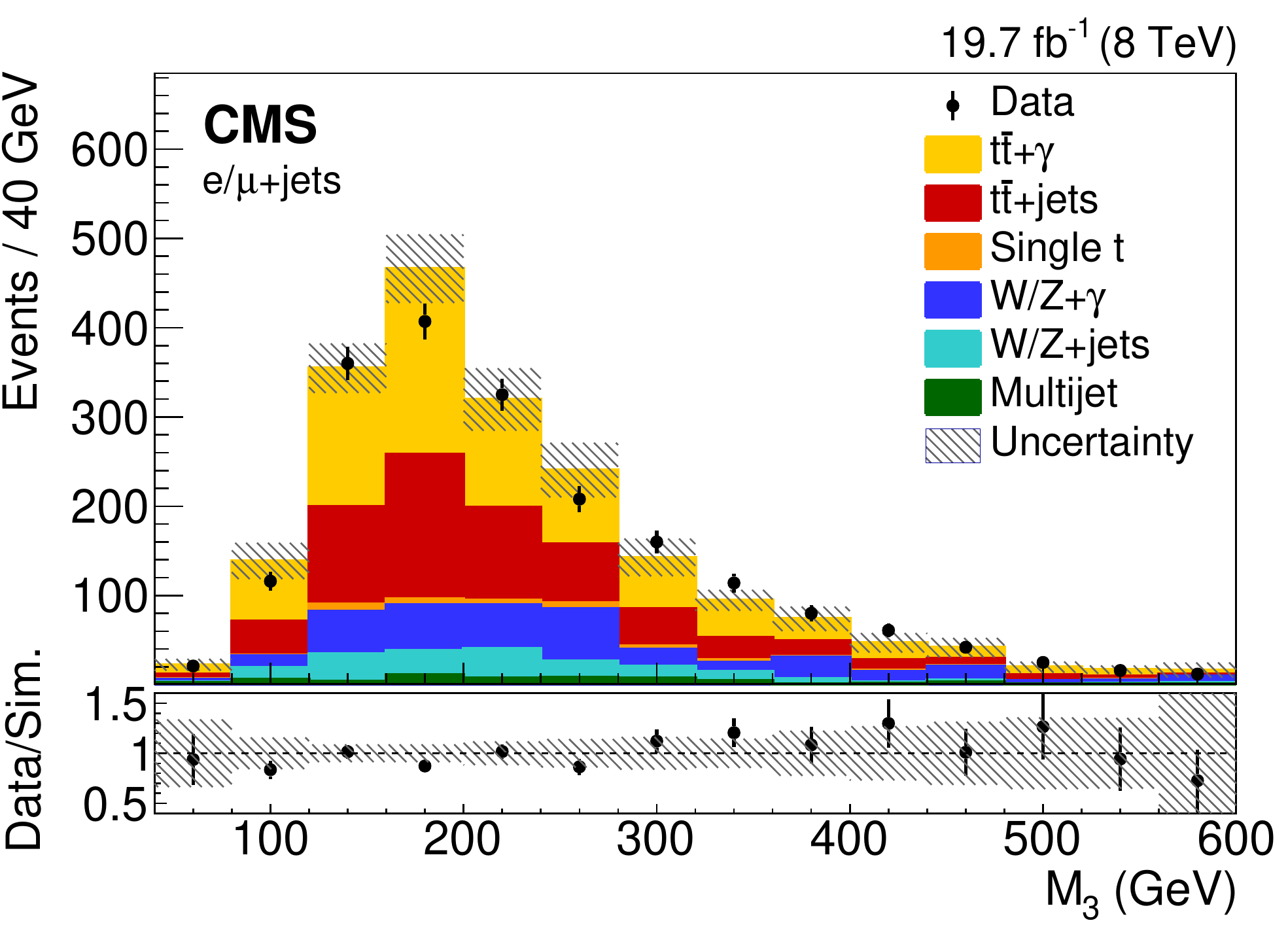}
      \caption{Distribution of the \Mthree variable in data and simulation, scaled to the result of the fit in a combination of the \ejets and \mujets channels, for events passing the photon selection.  The lower panel shows the ratio of the data to the prediction from simulation. The uncertainty band is a combination of statistical and systematic uncertainties in the simulation.}
      \label{fig:M3_photon}

  \end{figure}

Figure \ref{fig:M3_photon} shows the distribution of the \Mthree variable in data and simulation, scaled to the result of the fit.
From the fit result, the top quark purity is measured to be $\eJetsTopPurity \pm \eJetsTopPurityErr\stat$ and $\muJetsTopPurity \pm \muJetsTopPurityErr\stat$ in the \ejets and \mujets channels, respectively.
These are consistent with the expected values from simulation of $0.70\pm0.03\stat$ in the \ejets final state and $0.72\pm0.02\stat$ in the \mujets final state, where the uncertainties are due to the limited number of simulated events.

\section{Photon purity measurement} \label{sec:photonPurity}\label{sec:eToGammaSF}

Events are sorted into one of three categories based on the origin of their reconstructed photons.
Genuine photons are those which are promptly produced, originating from nonhadronic sources.
Misidentified photons can come from misreconstructed electrons, for which the track from the electron is not correctly reconstructed or properly matched to the energy cluster in the calorimeter, causing the electron to be reconstructed as a photon.
Quark or gluon fragmentation and hadronization processes can be misidentified as photons or yield genuine photons, which for both cases are expected to be nonisolated, in contrast with promptly produced photons.
The \ttgamma signal events predominantly fall within the first category while the latter two categories are mostly composed of background events.

Simulated events can be placed in one of these three categories based on matching between the reconstructed and generated photons.
Matching is performed based on the difference between the reconstructed photon and the generated particles in both \pt and the $\eta$-$\phi$ phase space.
If a reconstructed photon is matched to a generated photon from a nonhadronic source, it is classified in the first category.
Reconstructed photons that are not matched to a generated photon but instead are matched to a generated electron are classified as misidentified electrons, and placed in the second category.
All other events, which are not matched to either a generated photon or electron, are considered to be nonprompt photons originating from hadronic activity and placed in the third category.

Photons in the last category, which are produced from hadronic activity, are typically less isolated than genuine photons or misidentified electrons.
This difference in the isolation distribution is used to measure the photon purity, defined as the fraction of events with a photon originating from an isolated source (including both genuine photons and misidentified electrons).
A binned maximum-likelihood fit to the distribution of the charged-hadron isolation is used to measure the photon purity.

\begin{figure}
  \centering
    \includegraphics[width=0.48\textwidth]{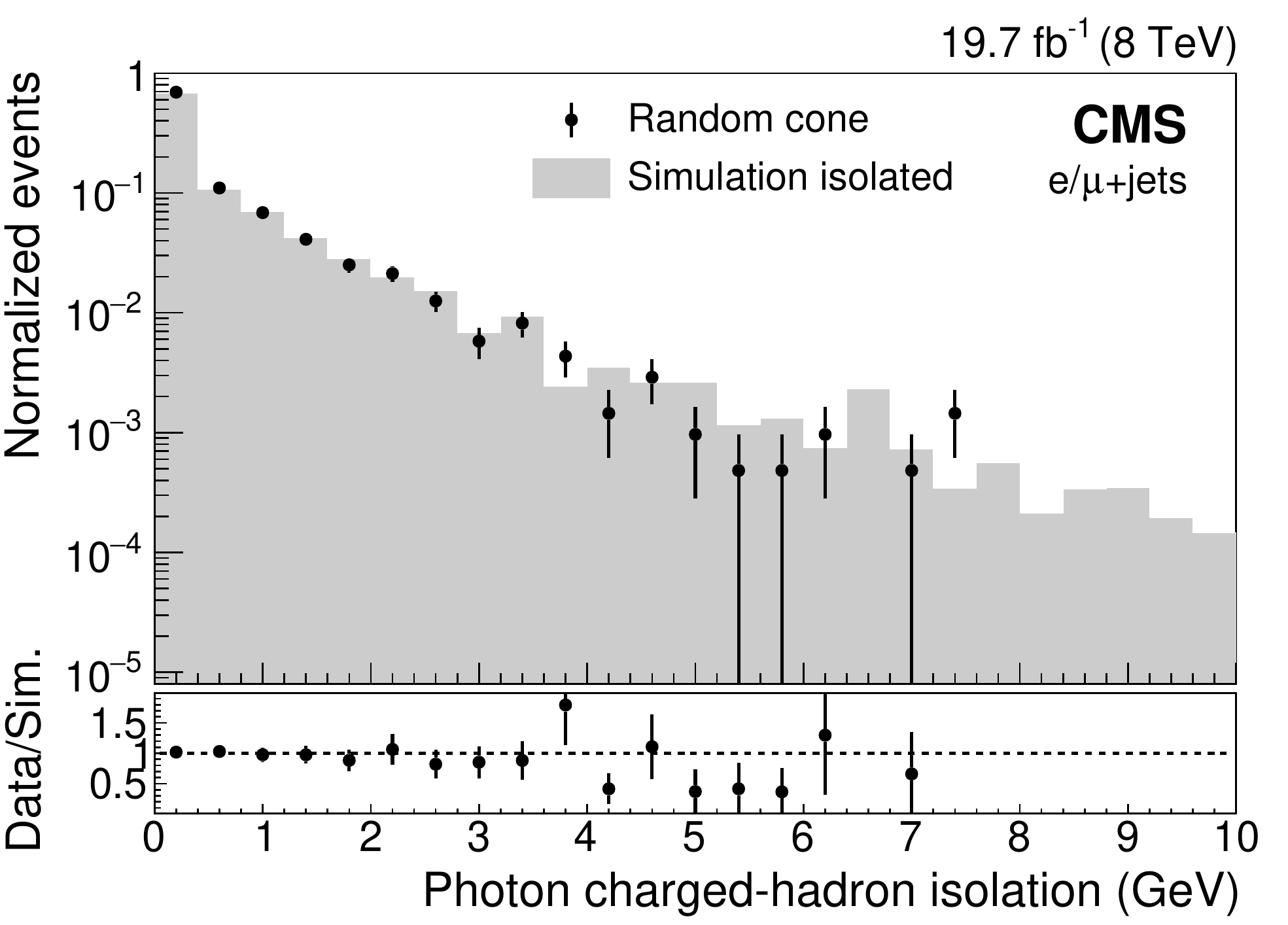}
    \includegraphics[width=0.48\textwidth]{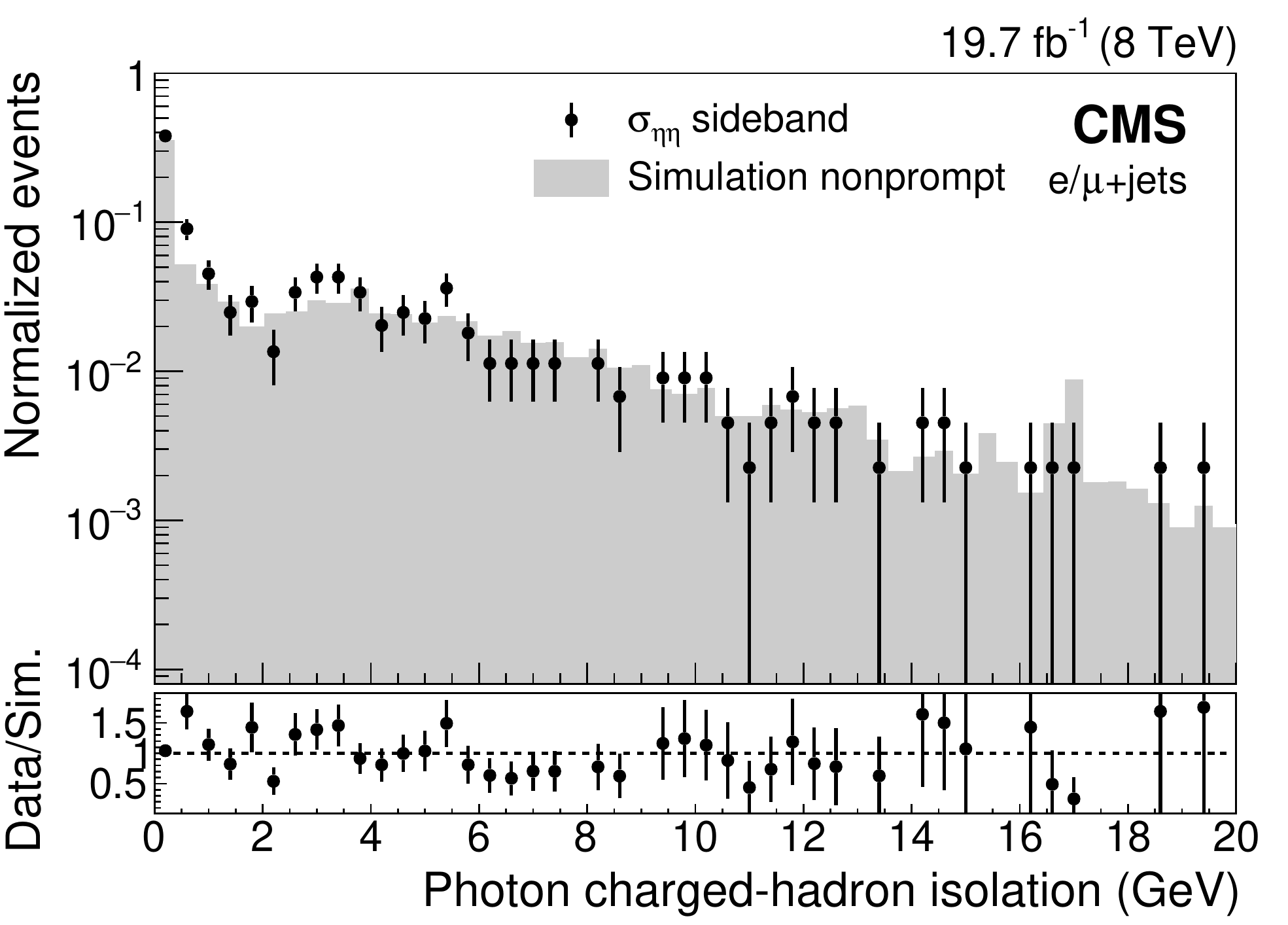}
    \caption{Shapes of isolated (left) and nonprompt (right) photon templates of the photon charged-hadron isolation, comparing templates derived from data to the distributions found from simulation in a combination of the \ejets and \mujets final states.  The lower panel shows the ratio of the distributions derived from data to those found from simulation.}
    \label{fig:ChHadSCFRtemplates}

\end{figure}

Templates for the shape of the charged-hadron isolation for isolated photons (coming from either genuine photons or misidentified electrons) and nonprompt (nonisolated) photons are taken from data.
The shape of the charged-hadron isolation for the isolated photon template is obtained using the random cone isolation method~\cite{CMSdiphoton}.
In this method, the sum of the transverse energy of PF charged-hadron candidates is measured within a cone of size $\Delta R = 0.3$ at the same $\eta$ value as the reconstructed photon, but in a random $\phi$ direction.
Contributions to the isolation sum from charged hadrons coming from pileup interactions are subtracted from the energy in the cone.
This gives an estimate of the isolation of a completely isolated particle.
The shape of the charged-hadron isolation for nonprompt photon events is taken from a sideband region.
The charged-hadron isolation of events with a photon having \sigmaietaieta between 0.012 and 0.016 is used to construct the template for nonisolated photons.
These events typically have nonprompt, hadronically produced photons.
Comparisons of the distributions of the charged-hadron isolation templates for isolated and nonprompt photons extracted from the data-based method and the templates taken from simulation using the generated particle matching are shown in Fig. \ref{fig:ChHadSCFRtemplates}.

\begin{figure}
  \centering
    \includegraphics[width=0.48\textwidth]{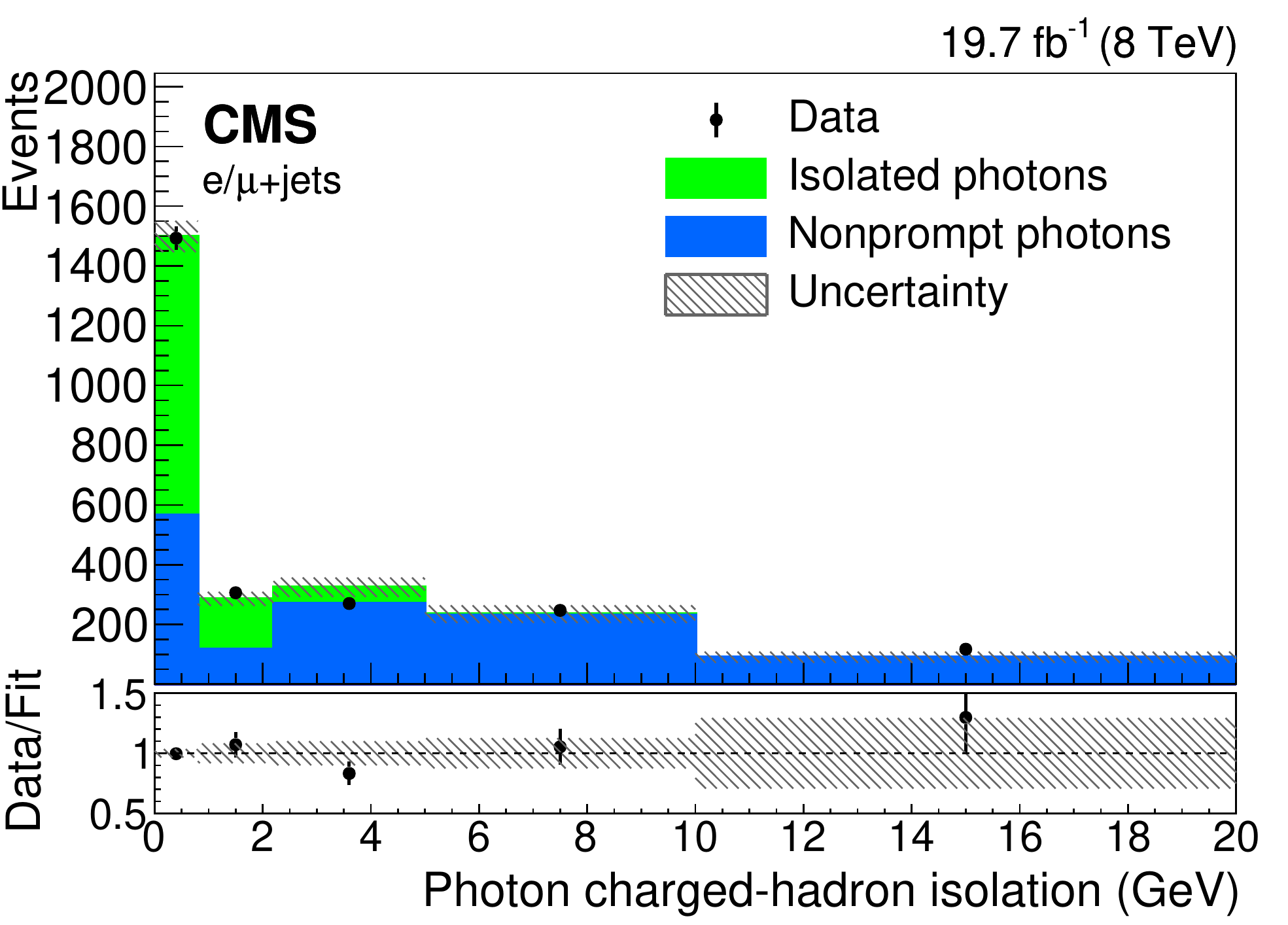}
    \caption{Result of the fit to the photon charged-hadron isolation in a combination of the \ejets and \mujets final states.  The uncertainty band shows the statistical uncertainties in the templates derived from data. The lower panel shows the ratio of the distribution observed in data to the sum of the templates scaled to the fit result.}
    \label{fig:ChHadSCFR_fit}

\end{figure}

In order to reduce the statistical fluctuations in the background template, the selection requirement of the photon charged-hadron isolation being less than 5\GeV is relaxed during the fit.
Instead, the fit is performed in the range of charged-hadron isolation less than  20\GeV, with all other photon selection requirements still in place.
The distribution suffers from lower statistical precision at higher values of the isolation, so the distribution is rebinned with larger bins for higher isolation values and finer binning for lower values where the statistical precision is better.
Figure \ref{fig:ChHadSCFR_fit} shows the result of the fit of the photon charged-hadron isolation in a combination of the \ejets and \mujets final state.
The photon purity is measured based on the fraction of events coming from isolated sources after the charged-hadron isolation requirement is put back in place.
The photon purity is measured to be $\eJetsPhoPurity \pm \eJetsPhoPurityErr\stat$ and $\muJetsPhoPurity \pm \muJetsPhoPurityErr\stat$ in the \ejets and \mujets final states, respectively.
The expected value for the photon purity in simulation, assuming the SM prediction for \ttgamma production, is $0.58 \pm 0.03$ in the \ejets final state and $0.57\pm0.02$ in the \mujets final state.

In order to correct the rate of misidentified electrons in simulation, the $\cPZ \to \Pep\Pem$ process is used to measure events in which one of the electrons from the $\cPZ$ boson decay is misidentified as a photon.
If the photon originates from a misidentified electron from the $\cPZ$ boson decay, the invariant mass of the combination of the electron and photon in the event will be near the $\cPZ$ boson mass.

Under the nominal event selection described previously, the contribution from $\cPZ$ boson production is highly suppressed and does not provide a large enough sample of events to measure the electron misidentification rate accurately.
In order to improve the statistical precision, the event selection is modified by relaxing the requirement of having a b-tagged jet in the event, while keeping all other requirements the same.
This enhances the contribution of $\cPZ\to \Pep\Pem$ events.
All steps for the multijet estimation and \Mthree fit are repeated for this new selection.

\begin{figure}
  \centering
    \includegraphics[width=0.48\textwidth]{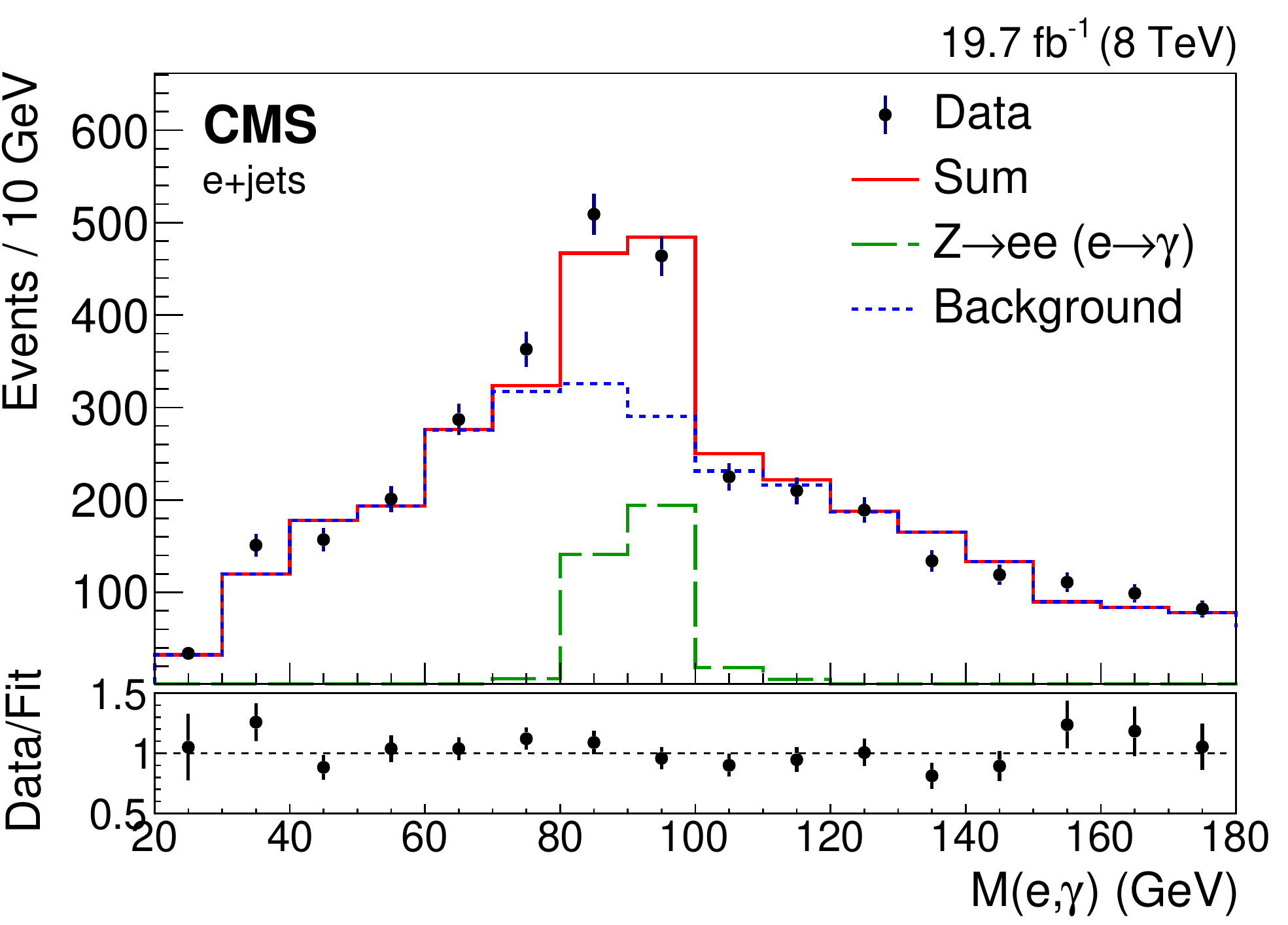}
    \caption{Result of the fit to the invariant mass of the electron and photon for events passing the modified event selection with the b tagging requirement relaxed.  Distributions are shown scaled to the results of the fit for $\cPZ \to \Pe\Pe \;(\Pe\to\gamma)$ and all other simulated samples (dashed lines), as well as the sum of the two samples (solid line).  The lower panel shows the ratio of the data to the simulation scaled to the fit results.}
    \label{fig:egammaFit}

\end{figure}

The removal of the b tagging requirement makes the $\cPZ$ boson mass peak much more pronounced in the $\Pe\gamma$ invariant mass distribution.
This allows a template fit to be performed, in order to estimate how well the misidentification of an electron as a photon is modeled in simulation.
The fit to the $\Pe\gamma$ invariant mass is performed using two templates, both derived from simulation.
The first template consists of events with $\cPZ$ bosons in which the reconstructed photon is matched to one of the electrons from the $\cPZ$ boson decay at the generator level.
The second template consists of all other simulated samples not included in the previous template and the data-based multijet sample.
The result of the fit is shown in Fig. \ref{fig:egammaFit}.
A scale factor of $\eGammaSF \pm \eGammaSFErr\stat$ is found for simulated events with a misidentified electron.
This scale factor is applied to all simulated events in which the photon is identified as originating from a misidentified electron.

\section[Top pair plus photon yield measurement]{The $\ttbar+\gamma$ yield measurement} \label{sec:likelihoodFit}

\begin{table}[htb]
\centering
\topcaption{Simulated samples categorized by reconstructed photon origin, after photon selection in the \ejets channel.
The data-based multijet sample is not expected to have signal photons or electrons.
All uncertainties combine statistical and systematic contributions.\label{tab:mcEventsAfterPhotonM3_ele}}
\begin{tabular}{l P{4.4} P{4.4} P{4.4} | P{4.4}}
\hline
Sample & \hd{Genuine photon} & \hd{Misid. electron} & \hdd{Nonprompt photon} & \hd{Total}\\
\hline
\ttgamma  & 312~17 & 0.2~0.1 & 8.5~0.9 & 321~17 \\
 \TTJets  & \hd{---} & 22~3 & 215~13 & 237~14 \\
 \Wgamma  & 75~25 & \hd{---} & \hdd{---} & 75~25 \\
 \WJets  & \hd{---} & \hd{---} & 60~15 & 60~15 \\
 \Zgamma  & 14~5 & 1.3~1.1 & \hdd{$0.5^{+0.7}_{-0.5}$} & 16~5 \\
 \ZJets  & \hd{---} & 43~28 & 11~6 & 54~30 \\
 Single t  & 11~3 & 2.0~1.3 & 16~4 & 29~7 \\
 QCD multijet  & \hd{---} & \hd{---} & 31~18 & 31~18 \\
 \hline
Total & 412~31 & 69~29 & 342~28 & 823~52 \\
Data & \hd{---} & \hd{---} & \hdd{---} & \hd{935} \\
\hline
\end{tabular}

\end{table}

\begin{table}[htb]
\centering
\topcaption{Simulated samples categorized by reconstructed photon origin, after photon selection in the \mujets channel.
The data-based multijet sample is not expected to have signal photons or electrons.
All uncertainties combine statistical and systematic contributions.\label{tab:mcEventsAfterPhotonM3_mu}}
\begin{tabular}{l P{4.4} P{4.4} P{4.4} | P{4.4} }
\hline
Sample & \hd{Genuine photon} & \hd{Misid. electron} & \hdd{Nonprompt photon} & \hd{Total} \\
\hline
\ttgamma  & 407~23 & 0.4~0.3 & 11~1 & 418~24 \\
 \TTJets  & \hd{---} & 31~5 & 291~16 & 322~17 \\
 \Wgamma  & 140~41 & \hd{---} & 9.0~6.7 & 149~45 \\
 \WJets  & \hd{---} & \hd{---} & 57~14 & 57~14 \\
 \Zgamma  & 21~7 & \hd{---} & 1.4~0.9 & 23~7 \\
 \ZJets  & \hd{---} & \hd{---} & 9.6~5.8 & 10~6 \\
 Single t  & 12~3 & 1.5~1.3 & 25~13 & 38~14 \\
 QCD multijet  & \hd{---} & \hd{---} & 36~20 & 36~20 \\
 \hline
Total & 580~48 & 33~5 & 440~33 & 1053~61 \\
Data & \hd{---} & \hd{---} & \hdd{---} & \hd{1136} \\
\hline
\end{tabular}

\end{table}

As previously mentioned, reconstructed photons originate from either a genuine photon, a misidentified electron, or a jet that produces a nonprompt photon.
Different processes contribute to each of these three categories in different ways.
For example, the \ttgamma and \Vgamma processes predominantly produce genuine photons, while the \ttbar and \VJets processes contribute to the nonprompt-photon or misidentified-electrons categories.
The breakdown of the number of events in the three reconstructed photon categories from each of the different simulated processes as well as the total number of expected and observed events are shown in Tables \ref{tab:mcEventsAfterPhotonM3_ele} and \ref{tab:mcEventsAfterPhotonM3_mu} for the \ejets and \mujets final states, respectively.

The modeling of misidentified electrons has been corrected using the scale factor described in Section \ref{sec:eToGammaSF}, but the modeling of nonprompt photons from jets remains uncorrected.
The normalization of the \TTJets, \WJets, \ZJets, and QCD samples have been cross-checked and corrected as described previously in Sections \ref{sec:QCDDYestimate} and \ref{sec:M3Fit}.
The contribution from single top quark processes is expected to be small and accurately modeled, and is left normalized to the theoretical  cross sections.
This leaves three major contributing sources, which have so far not been constrained and for which scale factors still need to be measured: \ttgamma, \Vgamma, and photons originating from jets.

The three remaining scale factors, the scale factor on \ttgamma simulation ($\ttgammaSF$), on \Vgamma simulation ($\VgammaSF$), and on simulation of photons originating from jets ($\jetToPhotonSF$), are derived by defining a likelihood function based on the three previously measured quantities: the photon purity, $\pi_{\Pe\gamma}^{\text{data}}$; top quark purity, $\pi_{\ttbar}^{\text{data}}$; and the number of events in data after the photon selection, $N^{\text{data}}$.
The likelihood function is defined as $\mathcal{L}(\ttgammaSF, \VgammaSF, \jetToPhotonSF) = \re^{- \chi^{2} / 2}$ where $\chi^{2}$ is the sum of three terms:
\begin{equation} \label{eqn:chi2}
\chi^{2}(\ttgammaSF, \VgammaSF, \jetToPhotonSF) =
\frac{ (\pi_{\Pe\gamma}^{\text{data}} - \pi_{\Pe\gamma}^{\text{MC}})^{2} }{ \sigma_{\pi_{\Pe\gamma}}^{2} } +
\frac{ (\pi_{\ttbar}^{\text{data}} - \pi_{\ttbar}^{\text{MC}})^{2} }{ \sigma_{\pi_{\ttbar}}^{2} } +
\frac{ (N^{\text{data}} - N^{\text{MC}})^{2} }{ \sigma_{N}^{2} },
\end{equation}
where $\pi_{\Pe\gamma}^{\text{MC}}, \pi_{\ttbar}^{\text{MC}}$, and $N^{\text{MC}}$ are the photon purity, top quark purity, and the number of events expected from simulation, and $\sigma_{\pi_{\Pe\gamma}}$ $\sigma_{\pi_{\ttbar}}$, and $\sigma_{N}$ are the statistical uncertainties in the measured quantities.
The value of the photon purity from simulation is taken to be the fraction of events in which the reconstructed photon originates from either a genuine photon or a misidentified electron.
Similarly, the top quark purity in simulation is found as the fraction of the total simulated events coming from either the \ttbar or \ttgamma processes.
Because these three values depend on the relative contribution of events from the different processes, they are functions of the three scale factors, $\ttgammaSF$, $\VgammaSF$, and $\jetToPhotonSF$.
For example, the photon purity would be increased for larger values of $\ttgammaSF$ or $\VgammaSF$ whereas $\jetToPhotonSF$ would increase the number of nonprompt photons and have the inverse effect on the photon purity.
Similarly the top quark purity would be increased for larger values of $\ttgammaSF$ or $\jetToPhotonSF$ (since \ttbar is the largest contributor of nonprompt photons), whereas $\VgammaSF$ has the inverse effect.
The likelihood fit is performed by scanning over the possible combinations of the three scale factors to find the one that results in values of $\pi_{\Pe\gamma}^{\text{MC}}, \pi_{\ttbar}^{\text{MC}},$ and $N^{\text{MC}}$, which most closely match the values observed in data, and thus returns the maximum likelihood value.

The likelihood fit is performed in the \ejets and \mujets final states individually, as well as in a combination of the two channels.
The combination is performed by maximizing the product of the likelihood functions from the \ejets and \mujets final states.

The scale factors obtained in the likelihood fit are applied to the simulation to extract the number of \ttgamma events observed, $N_{\ttgamma}$.
All \ttgamma events are scaled by \ttgammaSF, and those which fall within the nonprompt-photon category are additionally scaled by \jetToPhotonSF.
Applying the results of the fit in a combination of the \ejets and \mujets final states, $780 \pm 119\stat$ \ttgamma events are observed, $338 \pm 53\stat$ events and $442 \pm 69\stat$ events in the \ejets and \mujets final states, respectively.
The uncertainty comes predominantly from the statistical uncertainty in the results of the likelihood fit.

\section{Calculation of the cross section ratio} \label{sec:EffAcc}

The fiducial \ttgamma cross section ($\sigma^{\text{fid.}}_{\ttgamma}$) and the inclusive \ttgamma cross section ($\sigma_{\ttgamma}$) can be calculated based on the equations:
\begin{equation}
\sigma^{\text{fid.}}_{\ttgamma} = \frac{N_{\ttgamma}}{\epsilon^{\ttgamma} \, L}, \qquad \sigma_{\ttgamma} = \frac{N_{\ttgamma}}{A^{\ttgamma} \, \epsilon^{\ttgamma} \, L} = \frac{\sigma^{\text{fid.}}_{\ttgamma}}{A^{\ttgamma}},
\end{equation}
where $N_{\ttgamma}$ is the number of \ttgamma events observed, $A^{\ttgamma}$ is the acceptance of \ttgamma events within the fiducial phase space, $\epsilon^{\ttgamma}$ is the efficiency of the \ttgamma selection within events in the acceptance region, and $L$ is the integrated luminosity of the data set.

The acceptance is determined at generator level, by requiring generated events to fall within the fiducial phase space defined for the analysis.
Events are required to have exactly one generated prompt lepton in the fiducial phase space.
For electrons, this requires $\pt > 35$\GeV and $\abs{\eta} <2.5$ while not falling in the region $1.44 < \abs{\eta} < 1.56$.
The visible phase space for muons is defined as $\pt > 26$\GeV and $\abs{\eta} <2.5$.
Events are required to have at least three generated jets with $\pt > 30$\GeV and $\abs{\eta} <2.4$.
In order to replicate the \ptmiss requirement, the vector sum of the \pt of generated neutrinos is required to be greater than 20\GeV.
Lastly, events are required to have a generated photon with $\pt > 25$\GeV and $\abs{\eta}<1.44$.
The acceptance can be split into two components: the one coming from the branching fraction of \ttgamma to the \ejets or \mujets channels, and the one coming from the kinematic phase space requirements.
The kinematic acceptance is measured by the number of events passing the kinematic phase space requirements described above divided by the number of events generated in the \ejets and \mujets final states.

The efficiency is calculated as the ratio of reconstructed events that pass the event selection over the number of events generated in the fiducial phase space.
This accounts for the migration of events into and out of the fiducial phase space, and includes the efficiencies of the trigger requirement, object identification and reconstruction, and the event selection.
The measured values for the acceptance and efficiency of the \ttgamma selection in the \ejets and \mujets channels are given in Table \ref{tab:effAcc}.

\begin{table}[htb]
\centering
\topcaption{Kinematic acceptance and efficiency of the \ttgamma selection in the \ejets and \mujets final states. \label{tab:effAcc}}
\begin{tabular}{l c c}
\hline
 & \ejets & \mujets \\
\hline
Kinematic acceptance & $\eJetsKinAcc \pm \eJetsKinAccErr$ & $\muJetsKinAcc \pm \muJetsKinAccErr$ \\
Efficiency & $\eJetsFidEff \pm \eJetsFidEffErr$ & $\muJetsFidEff \pm \muJetsFidEffErr$ \\
\hline
\end{tabular}

\end{table}

In order to reduce the effect of systematic uncertainties that similarly affect all \TTJets production modes, the ratio of the cross section of fiducial \ttgamma production to the inclusive \ttbar production cross section is calculated as
\begin{equation}
R = \frac{ \sigma^{\text{fid}}_{\ttgamma} }{ \sigma_{\ttbar} } =
\frac{N_{\ttgamma}} {\epsilon^{\ttgamma}} \,
\frac{\epsilon^{\ttbar}_{\text{top}} A^{\ttbar}_{\text{top}} }{ N_{\ttbar}},
\end{equation}
where $N_{\ttbar}$ is the number of \ttbar events passing the top quark selection, and $\epsilon^{\ttbar}_{\text{top}}$ and $A^{\ttbar}_{\text{top}}$ are the efficiency and acceptance of top quark selection for \ttbar events.
The value of $\epsilon^{\ttbar}_{\text{top}} \, A^{\ttbar}_{\text{top}}$ is determined from simulation to be $\eJetsTTbarTopEffRounded$ in the \ejets final state and $\muJetsTTbarTopEffRounded$ in the \mujets final state with negligible statistical uncertainties.

\section{Sources of systematic uncertainty}

The effects of the systematic uncertainties are estimated by varying the simulated samples according to the uncertainty and repeating the measurement.
The top quark purity measurement, photon purity measurement, and likelihood fit are repeated for each source of systematic uncertainty and the new value of the cross section ratio is compared to the nominal value.
In this way, an estimate of the effect each source of systematic uncertainty has on the final result is found.
Table \ref{tab:systematics} lists the uncertainties in decreasing order of their effect on the cross section ratio, as found through the combination of the \ejets and \mujets final states.

The statistical uncertainty in the number of signal events found after maximizing the likelihood fit described in Section \ref{sec:likelihoodFit}, dominates the determination of the cross section for \ttgamma.
It includes the uncertainties in the measurement of the photon purity, top quark purity after photon selection, and the statistical uncertainty from the observed number of events in data.
The contribution from each of these three portions is estimated individually by performing the likelihood fit in which the uncertainties in these parameters are set to zero one at a time.
This effectively fixes the value to the measured value.
The change in the \ttgammaSF uncertainty (which is roughly 14\% in the standard likelihood fit) can be attributed to the fixed parameter.
The uncertainty is dominated by the top quark purity and photon purity uncertainties, which contribute 10\% and 9\%, respectively.
The statistical uncertainty caused by the limited number of events in data is approximately 4.8\%.

The uncertainties in the energy of reconstructed objects in the event are taken into account by scaling the energies of reconstructed objects in simulation up and down by the uncertainties in their corrections.
The uncertainties in the jet energy scale (JES) and jet energy resolution (JER)~\cite{JES} are applied to the reconstructed jets and the effect is propagated to the calculation of \ptmiss.
Similarly the uncertainty due to the photon energy is found by scaling the energy of reconstructed photons up and down by 1\%, and the measurement is repeated~\cite{CMS:EGM-14-001}.
The uncertainty due to the lepton energy scale is found by varying the \pt of the electrons and muons in the event by 1\% in the \ejets and \mujets final states, respectively~\cite{electronReconstruction,CMS-PAPER-MUO-10-004}.

A 50\% uncertainty is assigned to the normalization of the data-based multijet sample derived from the fit to the \ptmiss distributions.
Additionally, a 20\% normalization uncertainty is applied to the backgrounds that are fixed to their theoretical cross sections in the \Mthree fit (described in Section \ref{sec:M3Fit}).
The systematic uncertainty due to the scale factor for \ZJets simulation (described in Section \ref{sec:QCDDYestimate}) is applied by adjusting the scale factor up and down by its uncertainty.

The uncertainty in the efficiency of the b tagging algorithm is taken into account by varying the b tagging scale factors up and down by their uncertainties~\cite{BTV12001}.
Differences between the distribution of the \pt of the top quarks in data and simulation are taken into account by applying a reweighting based on the \pt of the generated top quarks and treating the difference from the nominal sample as a systematic uncertainty (``top quark \pt reweighting'')~\cite{CMSTopDifferential}.
The uncertainty in the pileup correction is found by recalculating the pileup distribution in data with a plus and minus 5\% change to the total inelastic proton-proton cross section~\cite{cmsInelasticPP}, and using these new distributions to reweight the simulation.

The uncertainty in the factorization and renormalization scales is taken into account by simulating the \ttgamma and \TTJets processes with the scales doubled and halved compared to the nominal value of $\mu_\mathrm{F} = \mu_\mathrm{R} = Q = \sqrt{\smash[b]{m^2_\cPqt + \Sigma \pt^2}}$ (where the sum is taken over all final state partons).
The uncertainty in the matching of partons at ME level to the parton shower (PS) is found by simulating \ttgamma and \TTJets processes with the threshold used for matching doubled and halved from the nominal value of 20\GeV.
The uncertainty arising from the choice of the top quark mass used in simulation is measured by simulating the samples with a value of $m_{\PQt}$ varied up and down by 1\GeV from its central value of $m_{\PQt}=172.5$\GeV.

\begin{table}[htb]
\centering
\topcaption{Uncertainties in the cross section ratio $R$ for the combination of the \ejets and \mujets final states.}
\label{tab:systematics}
\begin{tabular}{l | r }
\hline
Source & Uncertainty (\%) \\
\hline
Statistical likelihood fit   & 15.5  \\
Top quark mass   & 7.9  \\
JES   & 6.9  \\
Fact. and renorm. scale   & 6.7  \\
ME/PS matching threshold   & 3.9  \\
Photon energy scale   & 2.4  \\
JER   & 2.3  \\
Multijet estimate   & 2.0  \\
Electron misid. rate   & 1.3  \\
\ZJets scale factor   & 0.8  \\
Pileup   & 0.6  \\
Background normalization   & 0.6  \\
Top quark \pt reweighting   & 0.4  \\
b tagging scale factor   & 0.3  \\
Muon efficiency   & 0.3  \\
Electron efficiency   & 0.1  \\
PDFs & 0.1 \\
Muon energy scale   & 0.1  \\
Electron energy scale   & 0.1  \\
\hline
Total  &  20.7 \\
\hline
\end{tabular}

\end{table}

\section{Results}

The ratio of the fiducial cross section of \ttgamma to \ttbar production is found to be $R = \eJetsFidRatioWithError\statsyst$ in the \ejets final state and $R = \muJetsFidRatioWithError\statsyst$ in the \mujets final state.
The value of the fiducial \ttgamma cross section can be extracted from the cross section ratio using the measured \ttbar cross section of $244.9 \pm 1.4\stat^{+6.3}_{-5.5}\syst\pm 6.4\lum$\unit{pb}~\cite{cmsTTbarCXDilep2016}.
Multiplying the cross section ratio by the measured \ttbar cross section results in values for the \ttgamma fiducial cross section of $\eJetsFidCrossSecFB \pm \eJetsFidCrossSecFBErr\statsyst$\unit{fb} in the \ejets final state and $\muJetsFidCrossSecFB \pm \muJetsFidCrossSecFBErr\statsyst$\unit{fb} in the \mujets final state.

The value of the cross section times the branching fraction in the lepton+jets final states can be extrapolated from the fiducial cross section by dividing by the kinematic acceptance.
The kinematic acceptances (as given in Section \ref{sec:EffAcc}) are found to be $\eJetsKinAcc \pm \eJetsKinAccErr$ and $\muJetsKinAcc \pm \muJetsKinAccErr$ in the \ejets and \mujets final states.
This gives a cross section times branching fraction of $\sigma_{\ttgamma} \, \mathcal{B} = \eJetsExtrapolatedCX \pm \eJetsExtrapolatedCXErr$\unit{fb} in the \ejets final state and $\muJetsExtrapolatedCX \pm \muJetsExtrapolatedCXErr$\unit{fb} in the \mujets final state.
These values are in agreement with theoretical prediction of $592 \pm 71 (\text{scales}) \pm 30\,(\text{PDFs})$\unit{fb} for the cross section times branching fraction of each of the semileptonic final states~\cite{PhysRevD.83.074013}.

The combination of the \ejets and \mujets channels results in a cross section ratio per semileptonic final state of $\combinedFidRatioWithErrorPerChannel\statsyst$.
This results in a value of $\combinedFidCrossSecPerChannelFB \pm \combinedFidCrossSecErrPerChannelFB\statsyst$\unit{fb} for the fiducial \ttgamma cross section.
When extrapolated to the cross section times branching fraction by dividing by the kinematic acceptance, the result is $\sigma_{\ttgamma} \, \mathcal{B} = \combinedExtrapolatedCXPerChannel \pm \combinedExtrapolatedCXPerChannelErr$\unit{fb} per lepton+jets final state, in good agreement with the theoretical prediction.
Table \ref{tab:resultsSummary} summarizes the measured ratios and cross sections for the \ejets and \mujets final states as well as the combination.

\begin{table}[htb]
\centering
\topcaption{Cross section ratios, as well as fiducial and total cross sections per semileptonic final state.}
\label{tab:resultsSummary}
\begin{tabular}{l c c c}
\hline
Category & $R$ & $\sigma_{\ttgamma}^\text{fid}$ (fb) & $\sigma_{\ttgamma} \, \mathcal{B}$  (fb) \\
\hline
\ejets  & \eJetsFidRatioWithError & $\eJetsFidCrossSecFB\pm \eJetsFidCrossSecFBErr$ &  $\eJetsExtrapolatedCX\pm \eJetsExtrapolatedCXErr$ \\
\mujets  & \muJetsFidRatioWithError & $\muJetsFidCrossSecFB\pm \muJetsFidCrossSecFBErr$ &  $\muJetsExtrapolatedCX\pm \muJetsExtrapolatedCXErr$ \\
Combination  & \combinedFidRatioWithErrorPerChannel & $\combinedFidCrossSecPerChannelFB\pm \combinedFidCrossSecErrPerChannelFB$ &  $\combinedExtrapolatedCXPerChannel\pm \combinedExtrapolatedCXPerChannelErr$ \\
\hline
Theory & --- & --- & $592 \pm 71\ (\text{scales}) \pm 30\,(\text{PDFs})$ \\
\hline
\end{tabular}

\end{table}

\begin{figure}
    \centering
      \includegraphics[width=0.6\textwidth]{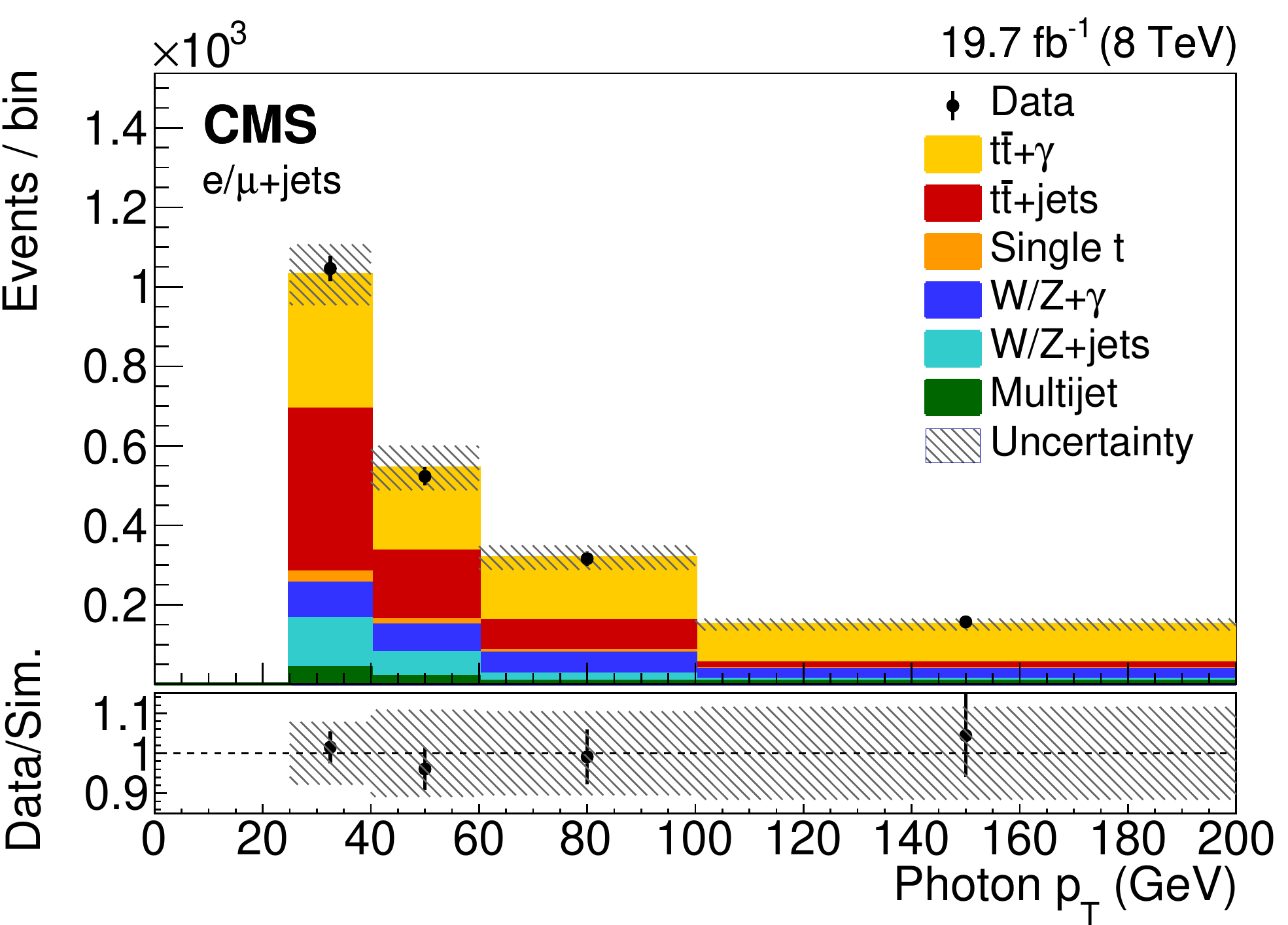}
      \caption{Distribution of the transverse momentum of the photon in data and simulation, scaled to the result of the likelihood fit in a combination of the \ejets and \mujets channels for events passing the photon selection.  The lower panel shows the ratio of the data to the prediction from simulation.  The uncertainty band is a combination of statistical and systematic uncertainties in the simulation.}
      \label{fig:Photon_Et}

  \end{figure}

\begin{figure}
    \centering
      \includegraphics[width=0.6\textwidth]{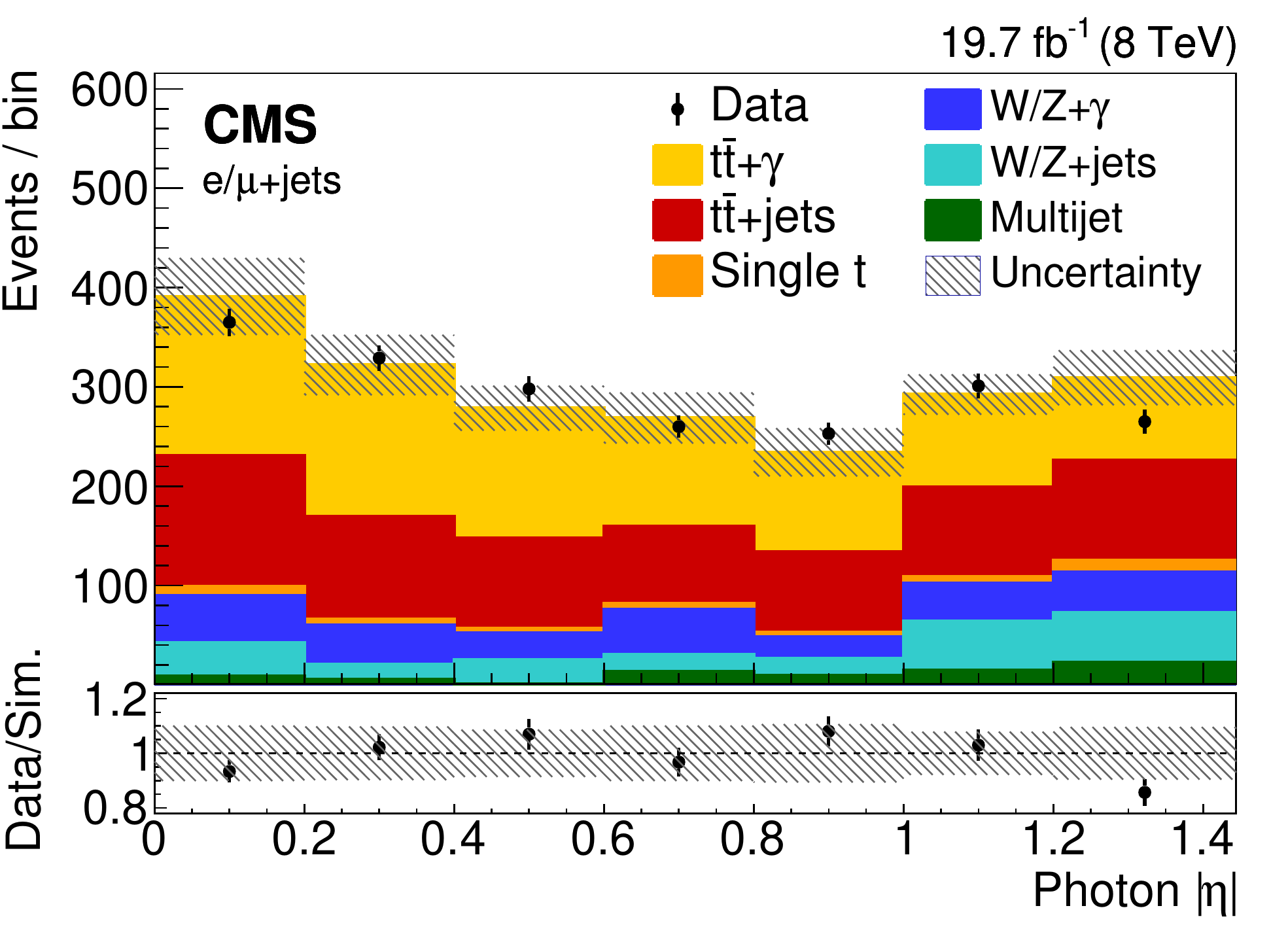}
      \caption{Distribution of the absolute value of the pseudorapidity of the photon in data and simulation, scaled to the result of the likelihood fit in a combination of the \ejets and \mujets channels for events passing the photon selection.  The lower panel shows the ratio of the data to the prediction from simulation.  The uncertainty band is a combination of statistical and systematic uncertainties in the simulation.}
      \label{fig:Photon_AbsEta}

  \end{figure}

The distributions of the transverse momentum and absolute value of the pseudorapidity of the photon candidate are shown in Figs. \ref{fig:Photon_Et} and \ref{fig:Photon_AbsEta}, scaled to the results of the likelihood fit.
While the statistical precision of this analysis currently limits the ability to perform a differential measurement of the \ttgamma cross section, there is the potential to measure the differential cross section in the future in both of these variables.

\section{Summary}

The results of a measurement of the production of a top quark-antiquark (\ttbar) pair produced in association with a photon have been presented.
The measurement is performed using 19.7\fbinv of data collected by the CMS detector at a center-of-mass energy of 8\TeV.
The analysis has been performed in the semileptonic \ejets and \mujets decay channels.

The ratio of the \ttgamma to \ttbar production cross sections has been measured to be $R={\sigma_{\ttgamma}/\sigma_{\ttbar}}={\combinedFidRatioWithErrorPerChannel}$.
By multiplying the measured ratio by the previously measured value of the \ttbar cross section, the fiducial cross section for \ttgamma production of $\combinedFidCrossSecPerChannelFB \pm \combinedFidCrossSecErrPerChannelFB$\unit{fb} has been found for events in the \ejets and \mujets final states.
The measured values are in agreement with the theoretical predictions.

\begin{acknowledgments}
We congratulate our colleagues in the CERN accelerator departments for the excellent performance of the LHC and thank the technical and administrative staffs at CERN and at other CMS institutes for their contributions to the success of the CMS effort. In addition, we gratefully acknowledge the computing centres and personnel of the Worldwide LHC Computing Grid for delivering so effectively the computing infrastructure essential to our analyses. Finally, we acknowledge the enduring support for the construction and operation of the LHC and the CMS detector provided by the following funding agencies: BMWFW and FWF (Austria); FNRS and FWO (Belgium); CNPq, CAPES, FAPERJ, and FAPESP (Brazil); MES (Bulgaria); CERN; CAS, MoST, and NSFC (China); COLCIENCIAS (Colombia); MSES and CSF (Croatia); RPF (Cyprus); SENESCYT (Ecuador); MoER, ERC IUT, and ERDF (Estonia); Academy of Finland, MEC, and HIP (Finland); CEA and CNRS/IN2P3 (France); BMBF, DFG, and HGF (Germany); GSRT (Greece); OTKA and NIH (Hungary); DAE and DST (India); IPM (Iran); SFI (Ireland); INFN (Italy); MSIP and NRF (Republic of Korea); LAS (Lithuania); MOE and UM (Malaysia); BUAP, CINVESTAV, CONACYT, LNS, SEP, and UASLP-FAI (Mexico); MBIE (New Zealand); PAEC (Pakistan); MSHE and NSC (Poland); FCT (Portugal); JINR (Dubna); MON, RosAtom, RAS, RFBR and RAEP (Russia); MESTD (Serbia); SEIDI, CPAN, PCTI and FEDER (Spain); Swiss Funding Agencies (Switzerland); MST (Taipei); ThEPCenter, IPST, STAR, and NSTDA (Thailand); TUBITAK and TAEK (Turkey); NASU and SFFR (Ukraine); STFC (United Kingdom); DOE and NSF (USA).

{\tolerance=1600
\hyphenation{Rachada-pisek} Individuals have received support from the Marie-Curie programme and the European Research Council and Horizon 2020 Grant, contract No. 675440 (European Union); the Leventis Foundation; the A. P. Sloan Foundation; the Alexander von Humboldt Foundation; the Belgian Federal Science Policy Office; the Fonds pour la Formation \`a la Recherche dans l'Industrie et dans l'Agriculture (FRIA-Belgium); the Agentschap voor Innovatie door Wetenschap en Technologie (IWT-Belgium); the Ministry of Education, Youth and Sports (MEYS) of the Czech Republic; the Council of Science and Industrial Research, India; the HOMING PLUS programme of the Foundation for Polish Science, cofinanced from European Union, Regional Development Fund, the Mobility Plus programme of the Ministry of Science and Higher Education, the National Science Center (Poland), contracts Harmonia 2014/14/M/ST2/00428, Opus 2014/13/B/ST2/02543, 2014/15/B/ST2/03998, and 2015/19/B/ST2/02861, Sonata-bis 2012/07/E/ST2/01406; the National Priorities Research Program by Qatar National Research Fund; the Programa Clar\'in-COFUND del Principado de Asturias; the Thalis and Aristeia programmes cofinanced by EU-ESF and the Greek NSRF; the Rachadapisek Sompot Fund for Postdoctoral Fellowship, Chulalongkorn University and the Chulalongkorn Academic into Its 2nd Century Project Advancement Project (Thailand); and the Welch Foundation, contract C-1845.
\par}
\end{acknowledgments}

\bibliography{auto_generated}
\cleardoublepage \appendix\section{The CMS Collaboration \label{app:collab}}\begin{sloppypar}\hyphenpenalty=5000\widowpenalty=500\clubpenalty=5000\textbf{Yerevan Physics Institute,  Yerevan,  Armenia}\\*[0pt]
A.M.~Sirunyan, A.~Tumasyan
\vskip\cmsinstskip
\textbf{Institut f\"{u}r Hochenergiephysik,  Wien,  Austria}\\*[0pt]
W.~Adam, E.~Asilar, T.~Bergauer, J.~Brandstetter, E.~Brondolin, M.~Dragicevic, J.~Er\"{o}, M.~Flechl, M.~Friedl, R.~Fr\"{u}hwirth\cmsAuthorMark{1}, V.M.~Ghete, C.~Hartl, N.~H\"{o}rmann, J.~Hrubec, M.~Jeitler\cmsAuthorMark{1}, A.~K\"{o}nig, I.~Kr\"{a}tschmer, D.~Liko, T.~Matsushita, I.~Mikulec, D.~Rabady, N.~Rad, B.~Rahbaran, H.~Rohringer, J.~Schieck\cmsAuthorMark{1}, J.~Strauss, W.~Waltenberger, C.-E.~Wulz\cmsAuthorMark{1}
% ...
\vskip\cmsinstskip
% ...
1:~~Also at Vienna University of Technology, Vienna, Austria\\
% ...

\end{sloppypar}
\end{document}